# Preliminary Simulation of Beam Extraction for the 28 GHz ECR Ion Source

**Bum-Sik Park*, Yonghwan Kim and Seokjin Choi**

*RISP, Institute for Basic Science, Daejeon 305-811, Korea*

The 28 GHz ECR(Electron Cyclotron Resonance) ion source is under development to supply various beams from proton to uranium at RISP(Rare Isotope Science Project). The superconducting magnet system for a 28 GHz ECR ion source consists of four solenoid coils and a saddle type sextupole. To meet the design requirement of ECR ion source, a numerical simulation was accomplished by using the KOBRA3-INP to optimize the extraction system which is the three dimensional ion optics code. The influence of the three dimensional magnetic field and the space charge effect was considered to extract the highly charged ion beam. In this paper, the design results of the extraction system were reported in detail.

Email: bspark@ibs.re.kr

Fax: +82-42-878-8866

# I. INTRODUCTION

The RISP (Rare Isotope Science Project) is under development of the heavy-ion accelerator facility to support various science programs [1]. The injector of the driver linear accelerator consists of a 28-GHz superconducting ECR ion source, the LEBT (Low Energy Beam Transport), the 500-keV/u RFQ (Radio Frequency Quadrupole) and the MEBT (Medium Energy Beam Transport).

The ECR ion source was designed to adopt superconducting magnets and dual high power RF sources of a 28 GHz gyrotron and an 18 GHz klystron to improve the performance. The magnetic design of the ion source has a mirror filed of 3.61 T at the injection and 2.07 T at the extraction side and a maximum radial field of 2.17 T. Therefore, the ion and the electron are magnetized by the strong axial magnetic field. So, it is important to know the initial properties of the extracted ion beam from the ECR ion source in order to achieve an effective transport and injection in the low-energy beamline. Because of the large emittance of the ion beams extracted from the ECR ion source, the low energy beam transport is more critical than the high energy beam transport of the extracted beam. And an ECR ion source possesses a large beam size and divergence generally because of the induced beam rotation due to the decreasing axial magnetic field [2]. Therefore, a numerical simulation has been conducted to optimize the extraction system including the electrode geometries. Through this kind of works, the emittance and the ion transmission were optimized. Additionally the information was can be obtained for the study of correcting the negative sextupole effect. This study investigated the full three dimensional simulation of ion beam extraction and transport from an ECR ion source.

# II. Extraction simulations

The KOBRA3D-INP code adopted to simulate ion beam sources and space-charge effects during beam extraction and transport [3]. In this code, the spatial distribution of the extracted ions is calculated by tracing ions from their starting positions along the magnetic field lines that pass through the extraction aperture. The main assumption is that the ions are magnetized in the whole plasma chamber and that the motion of the charged particles is determined by the magnetic field and space charge effect. Therefore the guiding center for the electrons and ions are strongly bound to the magnetic field lines. The major limitation of the code is that it does not simulate the production of multiply charged ions in the ECR ion source, because collisions and diffusion are not taken into account. Therefore, the code cannot calculate the charge state distribution (CSD) of the extracted ions. The charge state distribution should be acquired from the experimental observation or the calculation considering the particle balance equations.

Nevertheless, KOBRA3-INP can take into account the physical parameters: geometry, electric potential, magnetic flux densities and the space charge effect. To simulate the behavior of such an ion beam with an unknown space charge distribution, the following procedure was repeated until convergence is reached in the code. The iteration starts with computation of the electric potential on the system using the Laplace equation. The electric field is then calculated from the potential, and particles are traced through the system. The current carried by the particles is deposited onto the underlying mesh for the following solution of the Poisson equation for a new estimate of the electric potential. The iteration is continued like this until convergence is reached [4].

The magnetic field structure and the magnetic input data table for KOBRA3-INP were calculated by using the OPERA 3D program. Figure 1 shows the drawing of the ECR ion source. The plasma chamber, cryostat and double solenoid can be found. Figure 2 and 3 show the calculated magnetic field profile.

### III. Simulation Results

Table 1 shows the design requirements for the ECR ion source. To meet these requirements, the extraction system was designed by the 3D numerical simulation. In this simulation, the particles are started from the ECR zone which is defined by $w_c = eB/m$. For the 28GHz case, the surface, where the magnetic flux density is 1 T, is the ECR zone. The simulation is started with an equal number of Uranium 238 atoms (40000) and the charge state of +33.5 which are homogeneously distributed in the ECR zone. The initial conditions of these studies are summarized on table 2 which is on the basis of VENUS ECR ion source study [5].

The extraction of the $^{238}U^{33.5+}$ ions were studied by calculating their trajectories from ECR region to the plane behind the ground electrode of the extraction system. A configuration of the extraction region is shown in figure 4. The emittance was calculated according to the various extraction gaps as shown in figure 5. The plasma electrode is applied with +24.7 kV, the bias voltage is -800 V and the ground electrode connected to ground potential. The emittance is the smallest at the extraction gap is ~10mm. Figure 6 shows the calculated normalized rms emittance at the position behind 9 cm from the ground electrode according to the varying extraction voltages. On the basis of the simulation results, the extraction electrode gap was determined as 10mm and the extraction voltage will apply as 10 kV where the corresponding emittance is nearly satisfied the required emittance.

## IV. CONCLUSIONS AND FUTURE PLAN

The extraction system was designed by simulation study to reduce the emittance value which is generally huge due to the decreasing axial magnetic field. According to the calculation results by using the KOBRA3-INP, the required emittance of the RISP ECR ion source was accomplished. These results will be compared with the measurement results by using the pepper-pot emittance scanner and the Allison type emittance scanner.

## ACKNOWLEDGEMENT

This work was supported by the Rare Isotope Science Project of Institute for Basic Science funded by the Ministry of Science, ICT and Future Planning (MSIP) and the National Research Foundation (NRF) of Korea (2013M7A1A1075764).

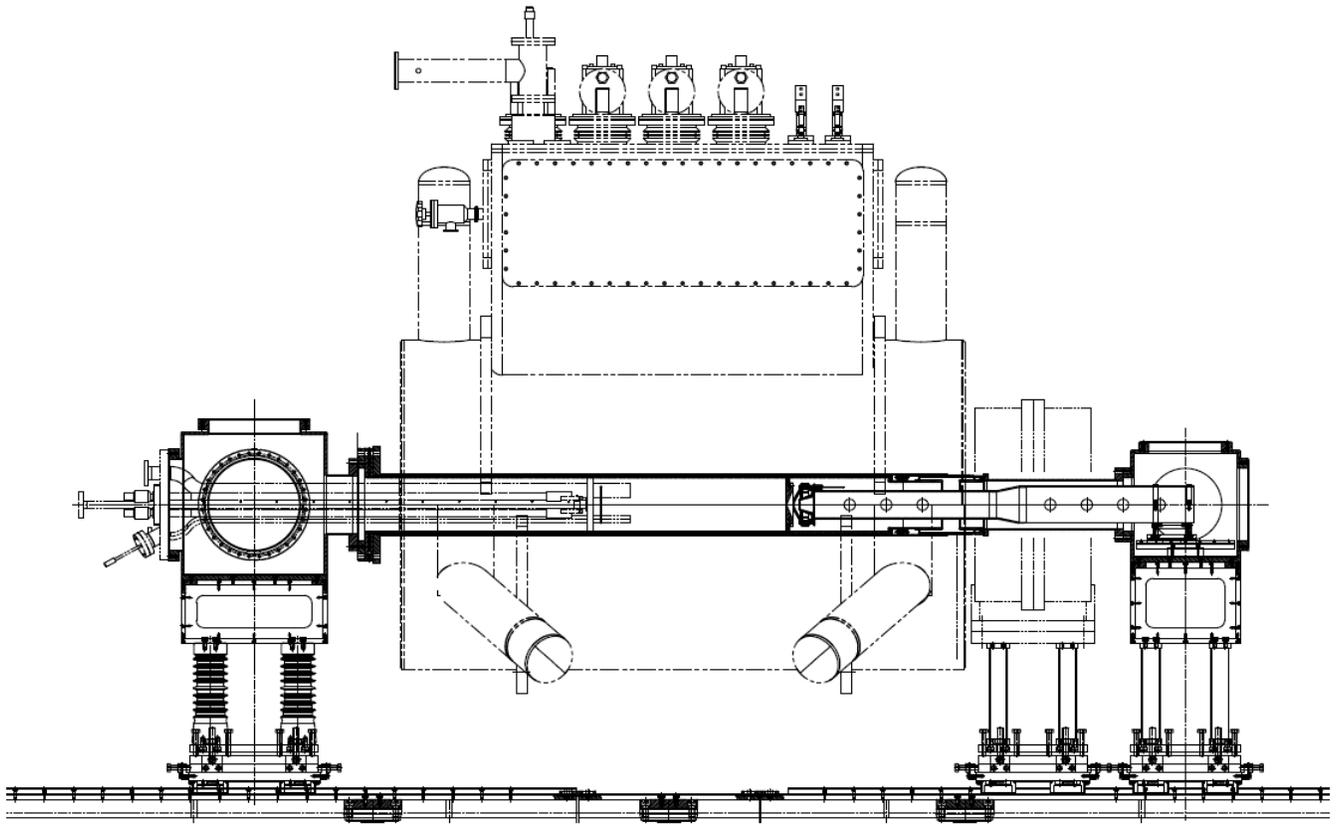

Figure 1. Mechanical layout of the ECR ion source including cryostat, plasma chamber and double solenoid. The plasma chamber was made with the aluminum. The superconducting coils are consisted with 4 solenoids and six saddle type hexapole coils. The inner diameter of the plasma chamber is 150mm and the mirror length is 500mm.

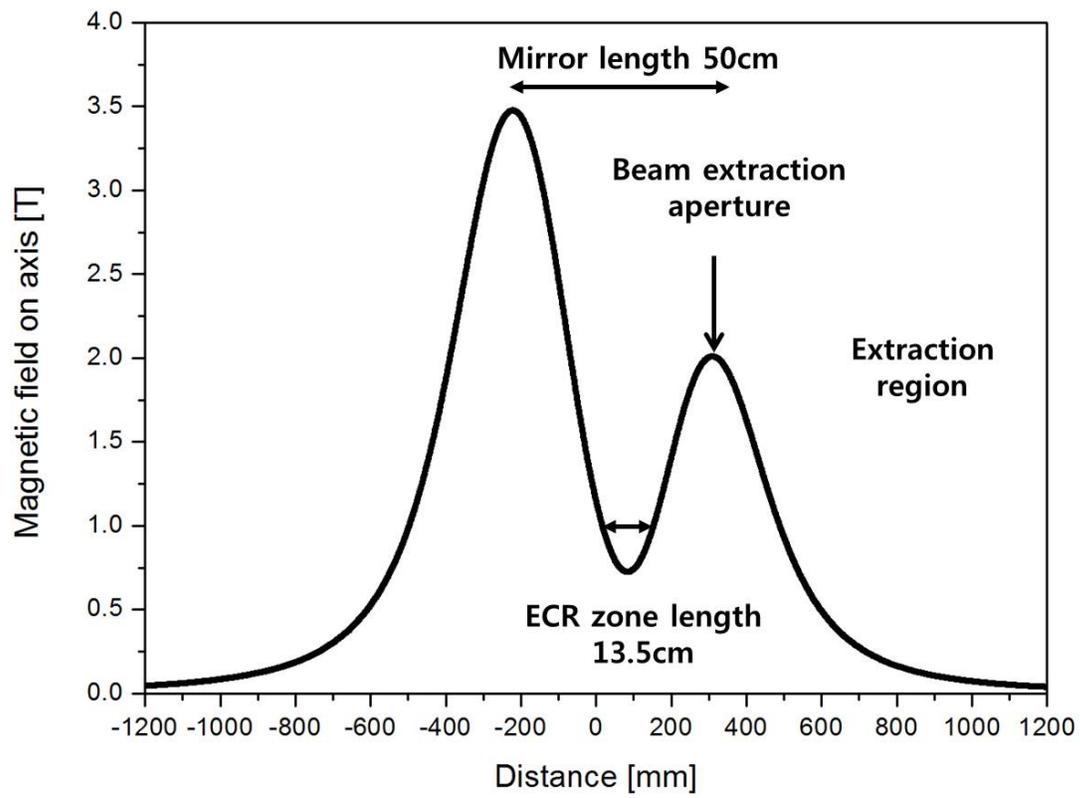

Figure 2. Axial magnetic field of the 28GHz ECR ion source according to the beam axis.

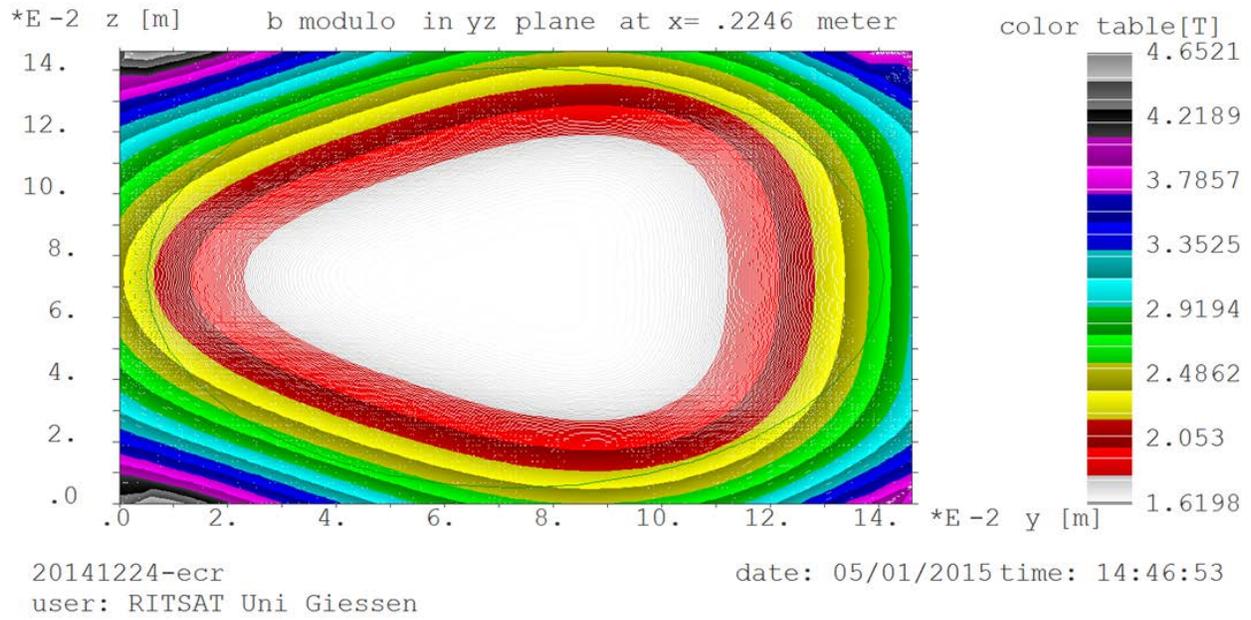

(a)

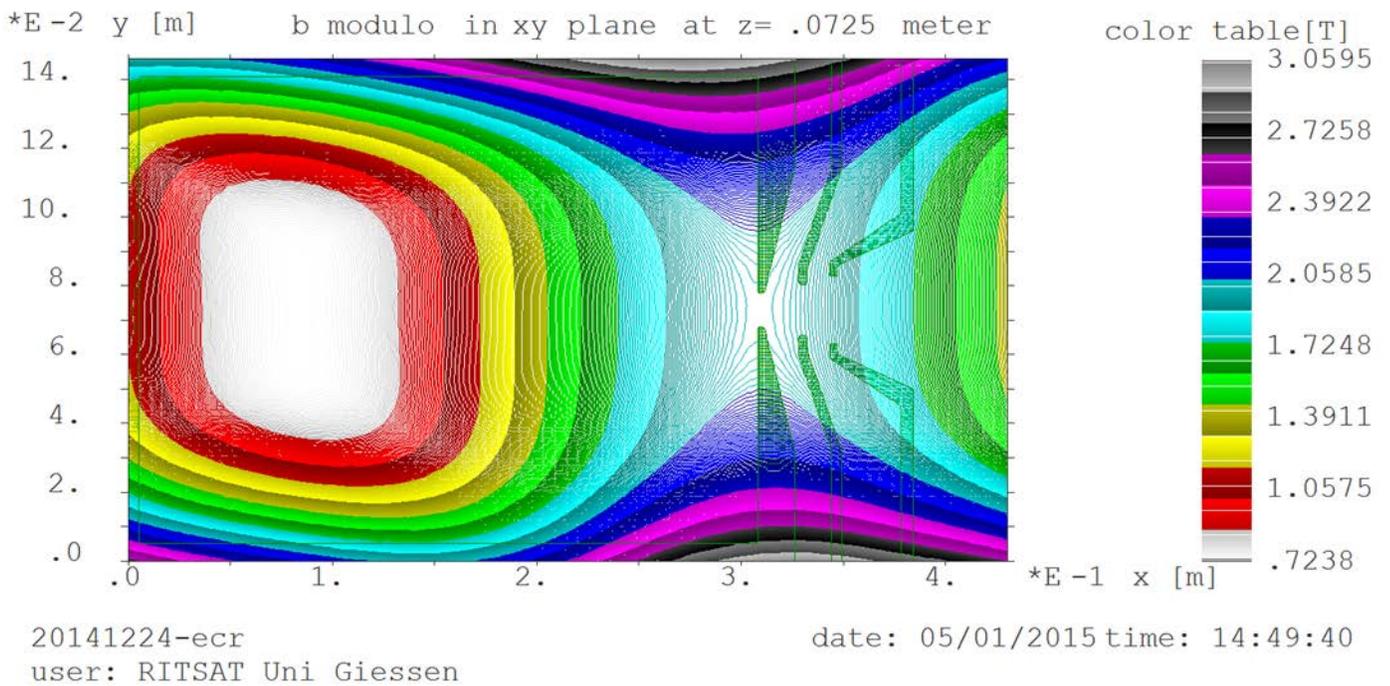

(b)

Figure 3. Magnetic field density distribution of ECR ion source at cross-section plane (a) and horizontal plane (b).

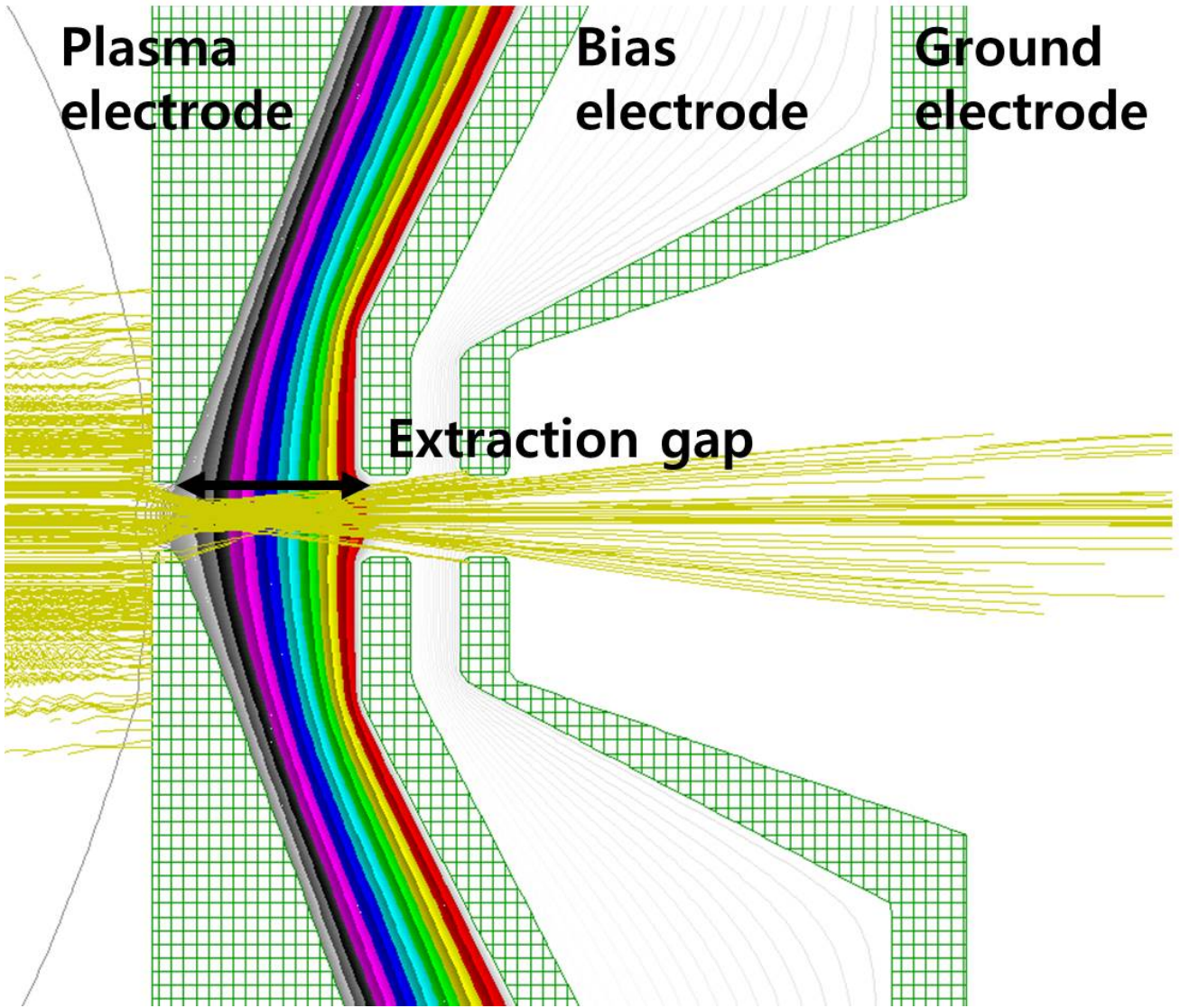

Figure 4. Configuration of the extraction area.

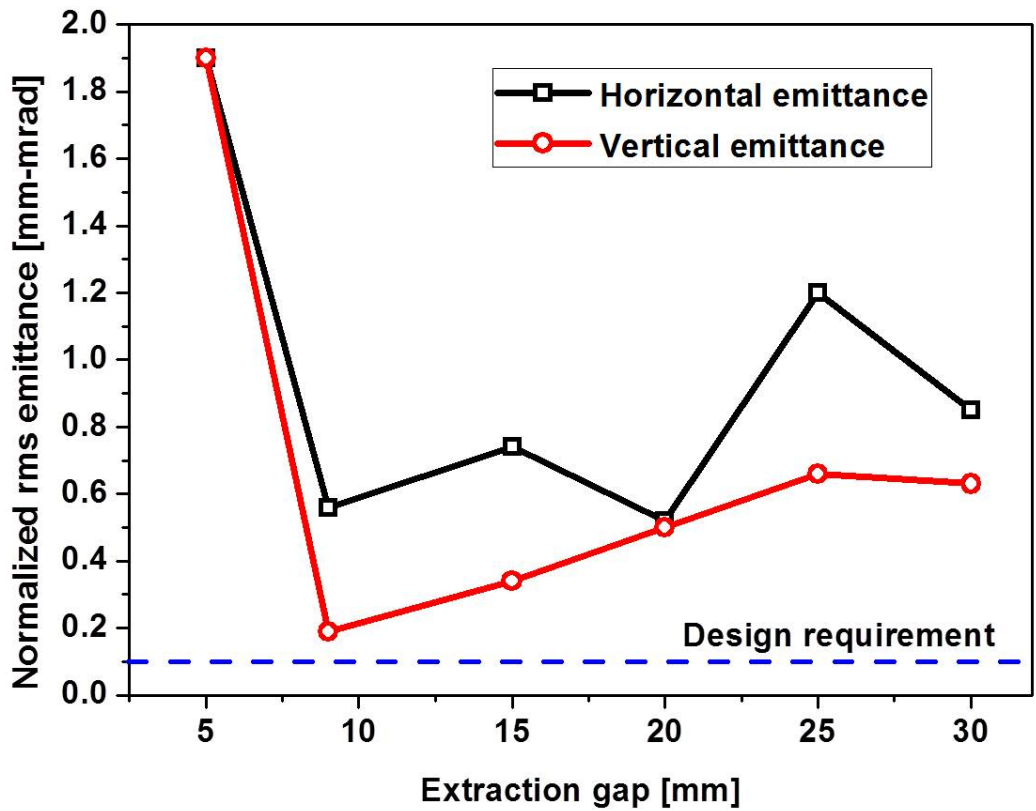

Figure 5. Simulated normalized rms emittance 9 cm behind the ground aperture of the extraction system according to the varying extraction gap. The plasma electrode is applied with +24.7 kV, the bias voltage is -800 V and the ground electrode connected to ground potential. The size of a plasma electrode hole is 5 mm in diameter

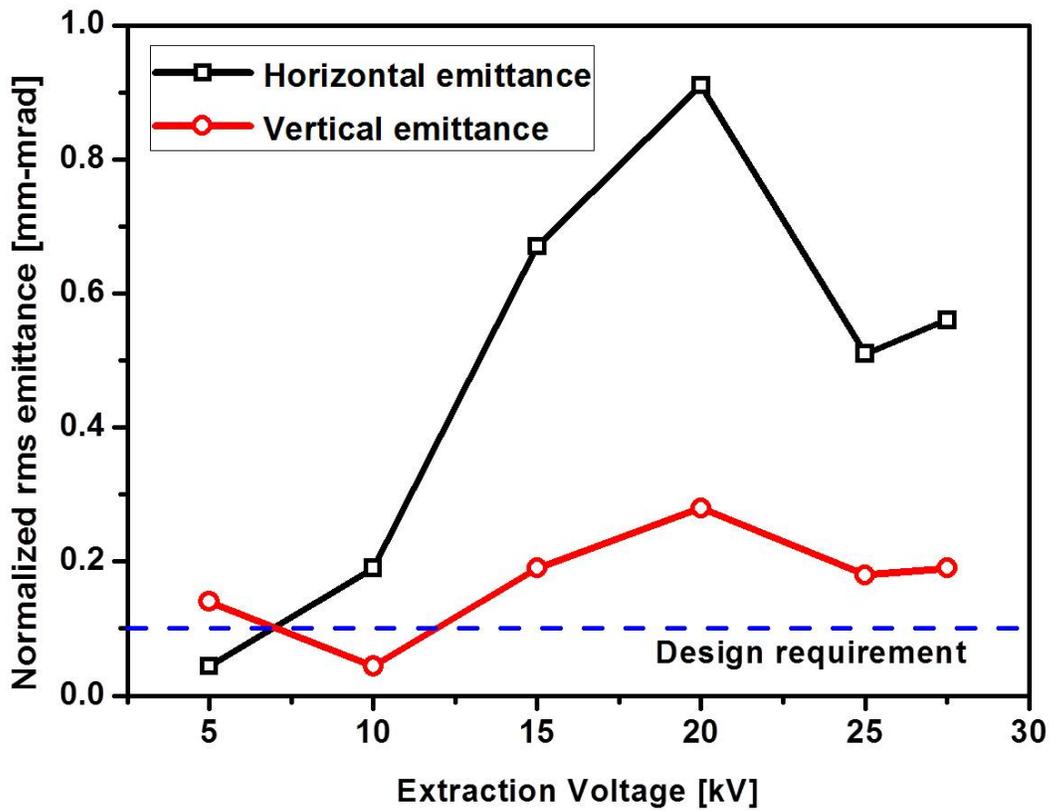

Figure 6. Simulated normalized rms emittance 9 cm behind the ground aperture of the extraction system according to the varying extraction voltage. The size of a plasma electrode hole is 5 mm in diameter and the extraction gap is 10 mm.

Table 1. Design parameters for the 28GHz ECR ion source

| Parameter | Value |
| --- | --- |
| Frequency | 28 + (18) GHz |
| RF Power | 10kW |
| Beam Current | 400euA for $^{238}U^{33+}$ + $^{238}U^{34+}$ |
| Plasma Chamber Material | Aluminum |
| Output norm(rms) emittance | 0.12 π mm-mrad |
| Output beam current | 10 keV/u |
| Number of Solenoid Coils | 4 |
| Sextupole Winding Type | Saddle type |

Table 2. Initial simulation parameters

| Parameter | Value |
| --- | --- |
| Ion mean temperature | 2eV |
| Kinetic energy | 3eV |
| Electron temperature | 5eV |
| Source voltage | 24.7kV |
| Bias voltage | -0.3kV |
| Total drain current | 12 p $\mu$A |
| Max. mag. Flux density at extraction | 2.1T |